\documentclass[twocolumn,tighten]{aastex62}

\usepackage{amsmath}
\usepackage{xspace}
\usepackage{multirow}
\usepackage{fancyapj}
\usepackage{lipsum}
\usepackage{soul}
\usepackage[normalem]{ulem}

\newcommand{\rxte}{\textit{RXTE}\xspace}

\newcommand{\nicer}{\textit{NICER}\xspace}
\newcommand{\exo}{\textit{EXOSAT}\xspace}
\newcommand{\source}{4U~1608--52\xspace}

\shortauthors{Jaisawal et al.}

\begin{document}

\title{\nicer observes a secondary peak in the decay of a thermonuclear burst from 4U~1608--52}

\author[0000-0002-6789-2723]{Gaurava K. Jaisawal}\email{gaurava@space.dtu.dk}
\affil{National Space Institute, Technical University of Denmark, 
  Elektrovej 327-328, DK-2800 Lyngby, Denmark}

\author[0000-0002-4397-8370]{J{\'e}r{\^o}me Chenevez}
\affil{National Space Institute, Technical University of Denmark, 
  Elektrovej 327-328, DK-2800 Lyngby, Denmark}

\author{Peter Bult}
\affiliation{Astrophysics Science Division, 
  NASA's Goddard Space Flight Center, Greenbelt, MD 20771, USA}

\author[0000-0002-4363-1756]{J.~J.~M.~in 't Zand}
\affil{SRON Netherlands Institute for Space Research, Sorbonnelaan 2, 3584 CA Utrecht, The Netherlands}

\author[0000-0002-6558-5121]{Duncan K.~Galloway}
\affiliation{School of Physics \& Astronomy, Monash University, Clayton VIC 3800, Australia}
\affiliation{also Monash Centre for Astrophysics, Monash University}

\author[0000-0001-7681-5845]{Tod E. Strohmayer} 
\affil{Astrophysics Science Division and Joint Space-Science Institute,
 NASA's Goddard Space Flight Center, Greenbelt, MD 20771, USA}

\author[0000-0002-3531-9842]{Tolga G{\"u}ver}
\affiliation{Department of Astronomy and Space Sciences, Science Faculty, 
Istanbul University, Beyaz{\i}t, 34119 Istanbul, Turkey}
\affiliation{Istanbul University Observatory Research and Application Center, Beyaz{\i}t, 34119 Istanbul, Turkey}

\author{Phillip Adkins} 
\affiliation{Safety Division, 
  NASA's Goddard Space Flight Center, Greenbelt, MD 20771, USA}

\author[0000-0002-3422-0074]{Diego Altamirano}
\affiliation{Physics \& Astronomy, University of Southampton, 
  Southampton, Hampshire SO17 1BJ, UK}

\author{Zaven Arzoumanian} 
\affiliation{Astrophysics Science Division, 
  NASA's Goddard Space Flight Center, Greenbelt, MD 20771, USA}

\author[0000-0001-8804-8946]{Deepto Chakrabarty}
\affil{MIT Kavli Institute for Astrophysics and Space Research, 
  Massachusetts Institute of Technology, Cambridge, MA 02139, USA}

\author{Jonathan Coopersmith} 
\affiliation{KBRwyle and Quality \& Reliability Division, 
  NASA's Goddard Space Flight Center, Greenbelt, MD 20771, USA}

\author{Keith C. Gendreau} 
\affiliation{Astrophysics Science Division, 
  NASA's Goddard Space Flight Center, Greenbelt, MD 20771, USA}

\author[0000-0002-6449-106X]{Sebastien Guillot} 
\affil{CNRS, IRAP, 9 avenue du Colonel Roche, BP 44346, F-31028 Toulouse Cedex 4, France} 
\affil{Universit\'e de Toulouse, CNES, UPS-OMP, F-31028 Toulouse, France} 

\author{Laurens Keek}
\affil{Department of Astronomy, University of Maryland, College Park, MD 20742, USA}

\author[0000-0002-8961-939X]{Renee M. Ludlam}
\affil{Department of Astronomy, University of Michigan, 
  1085 South University Ave, Ann Arbor, MI 48109-1107, USA}

\author{Christian Malacaria}
\affiliation{NASA Marshall Space Flight Center, NSSTC, 320 Sparkman Drive, Huntsville, AL 35805, USA}\thanks{NASA Postdoctoral Fellow}
\affiliation{Universities Space Research Association, NSSTC, 320 Sparkman Drive, Huntsville, AL 35805, USA}

\begin{abstract}

We report for the first time below 1.5 keV, the detection of a  secondary peak in an Eddington-limited thermonuclear X-ray burst observed by the {\it Neutron Star Interior Composition Explorer} (\nicer) from the low-mass X-ray binary 4U~1608--52. Our time-resolved spectroscopy of the burst is consistent with a model consisting of a varying-temperature blackbody, and an evolving persistent flux contribution, likely attributed to the accretion process. The dip in the burst intensity before the secondary peak is also visible in the bolometric flux.  Prior to the dip, the blackbody temperature reached a maximum of $\approx3$~keV. Our analysis suggests that the dip and secondary peak are not related to photospheric expansion, varying circumstellar absorption, or scattering. Instead, we discuss the observation in the context of hydrodynamical instabilities, thermonuclear flame spreading models, and re-burning in the cooling tail of the burst.

\end{abstract}


\section{Introduction} \label{sec:intro}

Thermonuclear (type I) X-ray bursts originate in the unstable 
burning of hydrogen- or helium-rich material on the surface of 
a neutron star (for reviews see \citealt{Lewin1993, Strohmayer2003, 
Galloway2017}). This material is typically accreted from a (sub-) 
solar mass companion through Roche-lobe overflow in low mass X-ray 
binaries (LMXBs). Type-I X-ray bursts (simply bursts hereafter) are 
characterized by a few-second rise in X-ray luminosity by at least an order 
of magnitude and lasting tens to hundreds of seconds. Their X-ray emission 
during the decaying part of the burst is consistent with a cooling blackbody 
with a 2--3~keV peak temperature. It is commonly assumed that the 
``persistent'' emission from the accretion process remains constant
during the burst.   However, recent studies suggest 
that the irradiation from bursts can modify 
the persistent continuum (\citealt{Chen2012, 
Zand2013, Worpel2013, Worpel2015,  Degenaar2018, Keek2018}).  
These effects can be interpreted as reprocessing/reflection from the disk 
\citep{Ballantyne2004}, changes in the accretion flow rate through 
Poynting-Robertson drag \citep{Walker1992}, or cooling of the corona 
\citep{Ji2014a}.

The most luminous bursts reach the Eddington limit: the outward 
radiation pressure overcomes the gravitational binding energy, leading to 
photospheric radius expansion (PRE; \citealt{Ebisuzaki1983, Lewin1984}). 
PRE bursts show a sudden drop in temperature and an increase 
in the photospheric radius by tens of kilometers above the surface 
(\citealt{Kuulkers2003, Keek1820}). As it expands, the photosphere cools 
causing its thermal spectrum to shift to lower energies, and possibly 
out of the passband of hard X-ray instruments such as those flown on 
the {\it Rossi X-ray Timing Explorer} (\rxte) and {\it INTEGRAL}. 
This spectral shift causes a drop in the measured intensity for these 
instruments. Following its expansion phase, the photosphere falls back 
onto or close to the neutron star surface, heats up, 
and its thermal spectrum therefore re-enters the hard X-ray band, 
causing a secondary increase of the measured intensity. This passband 
limitation of hard X-ray instruments is usually responsible for the 
double-peaked structure observed, while the bolometric flux is 
single-peaked  (\citealt{Fujimoto1989, Galloway2008}).

Some sources such as  4U~1636--536 \citep{Bhattacharyya2006} and GX~17+2 
\citep{Kuulkers2002} are known to show intrinsically double-peaked bursts. 
Despite the fact that these events were non-PRE bursts, their bolometric 
flux contained a dip-like structure. The Rapid Burster is another example 
where six double-peaked type-I (non-PRE) bursts have been detected during the 
soft to hard state transition \citep{Bagnoli2014}. A very rare triple-peaked 
burst is also known from  4U~1636--536 \citep{Zhang2009}.

In this  paper, we study the flux and spectral evolution of a burst that 
included a second peak during its cooling tail, as observed from 
the atoll source \source using the 
{\it Neutron Star Interior Composition Explorer} (\nicer, 
\citealt{Gendreau2016, Gendreau2017}).

The X-ray burster \source is a well-known transient LMXB that was discovered 
in 1971 with two {\it Vela-5} satellites (\citealt{Belian1976, Grindlay1976, 
Tananbaum1976}). The neutron star in the system accretes from the late F or 
early G-type star QX Nor---a source rich in hydrogen and helium---in an 
orbit of period 0.537~d (\citealt{Grindlay1978, Wachter2002}).

4U 1608--52 moves through different accretion states during an outburst. 
At low luminosities, the source is in the so-called hard spectral state, 
where its spectrum is dominated by a hard power-law component. At higher luminosities, 
the accretion-disk transitions to the soft state and exhibits a spectrum dominated by 
soft thermal photons (see, e.g. \citealt{Done2007} for spectral state classification). 
\source usually shows bursts in both the soft 
(banana branch) and hard (island) states. \citet{Ji2014} showed that 
the bursts affect the persistent emission differently based on the 
spectral state. The persistent flux observed in the soft state increases 
across the burst, while this behaviour holds in the hard state only 
when the burst is non-PRE. A decreasing persistent flux  
is observed for brighter events in the hard state  \citep{Ji2014}. 

Thanks to observations of PRE bursts, the source distance is known to lie within 
the range 2.9--4.5~kpc (\citealt{Galloway2008, Guver2010}). The spin period 
is constrained to $\approx620$~Hz based on the detection of burst 
oscillations (\citealt{Muno2001, Galloway2008}). Other physical parameters 
of the neutron star determined from burst time-resolved spectroscopy are
a mass $M=1.2$--1.6~$M_\odot$ and a radius of 13 to 16~km  
(\citealt{Poutanen2014}; see also \citealt{Ozel2016}). 
In addition to regular bursts, one superburst, likely due to the deep burning 
of a thick carbon layer, was observed in 2005 \citep{Keek2008}.

 Using high timing and spectral capabilities of \nicer in soft X-rays, we study a 
double-peaked burst from \source for the first time below 1.5~keV.  The present paper 
focuses on the nature of this  event and also examines the effect of the burst emission 
on the accretion environment using a variable persistent flux method. We describe the observations and our analysis 
methods in Section~\ref{sec:obs}, and present our results and discussion in 
Sections~\ref{sec:results} and \ref{sec:dis}, respectively.

\begin{figure}
\centering
\includegraphics[height=3.35in, width=2.72in, angle=-90]{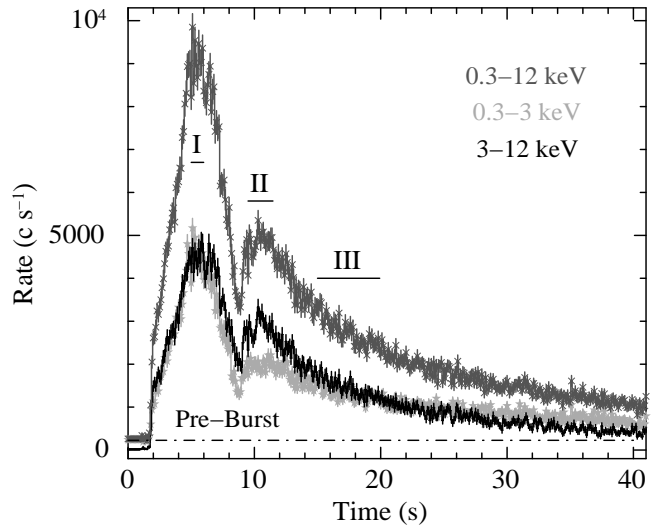}
\caption{Burst light curve observed with \nicer at 0.1~s resolution. 
A re-brightening is detected at all energies $\simeq$5~s after the 
primary peak. The pre-burst count rate (horizontal line) is $\sim$226~c~s$^{-1}$ 
in the 0.3--12 keV band. The segments I, II, and III represent broad time
spans used for time-resolved spectroscopy at the first peak, 
second peak, and in the decay part of the burst,  
respectively.} 
\label{lc}
\end{figure}

\section{Observations and Analysis} \label{sec:obs}

Launched in June 2017, the \nicer X-ray Timing Instrument 
(XTI, \citealt{Gendreau2016}) is a non-imaging soft X-ray 
telescope attached to the International Space Station. It 
consists of 56 co-aligned concentrator optics, each paired 
with a silicon-drift detector  \citep{Prigozhin2012}. This 
instrument records photons between 0.3--12~keV at an 
unprecedented time resolution of $\approx$100~ns  and spectral  
resolution of $\approx$100~eV (full width at half maximum). The peak 
effective area of the 52 
currently active detectors is $\approx$1900~cm$^2$ at $1.5$~keV.

\nicer{} monitored the transient source \source actively as part of 
the mission's baseline science program. Only two type-I bursts have been 
observed (in publicly available data sets  ObsID 0050070101--0050070110, 
1050070101--1050070174, and 2050070101 --2050070111) over a net exposure of 
180.3~ks in between 2017 June and 2019 April. The 
first burst was  observed on 2017 June 25 (MJD~57929.5002, ObsID: 
0050070102), reaching a peak intensity of 6230$\pm$250~c~s$^{-1}$ in the 0.3--12~keV 
band, whereas the second was detected on 2017 September 28 
(MJD~58024.2294, ObsID: 1050070103), peaking at 9840$\pm$306~c~s$^{-1}$. 
The latter event is the focus of the present study.  

We processed the data using \textsc{heasoft} version 6.24, 
\textsc{nicerdas} version 2018-04-24\_{V004} and the calibration 
database version 20180711. Good time intervals  (GTIs) were created 
via {\tt NIMAKETIME} using the standard filtering criteria.  
We applied these GTIs on 
processed XTI data to produce the spectra and lightcurves. 
For the spectral study, we used \texttt{XSPEC} version 12.10.0 \citep{Arnaud1996} 
along with \nicer response and effective area 
files version 1.02. The background contribution to our 
observations is determined from \nicer observations of an 
\rxte blank-sky region ($\sim$1--2 c~s$^{-1}$ from \rxte-6; 
\citealt{Jahoda2006}).

\section{Results} \label{sec:results}

\subsection{Burst Light Curve} \label{sec:lc}

Figure~\ref{lc} shows 0.1-s binned lightcurves of the burst
detected by \nicer\ on 2017 September 28 (MJD 58024.2294), 
in the 0.3--12 keV, 0.3--3 keV, and 3--12 keV bands.
The intensity remained above 20\% of the peak count rate for 
$\simeq$20~s. The burst reached a maximum count rate of 
9840 c~s$^{-1}$ (0.3--12 keV) 3.5~s after onset. 
About 5~s later, after a dip, a second peak at 5200~c~s$^{-1}$ 
occurred in the burst tail (Figure~\ref{lc}). The observed count 
rate of the first  peak was nearly the same in the 0.3--3 and 
3--12 keV energy bands, while the second peak was comparatively 
fainter in the soft ($\le$3~keV) X-rays. The dip observed between 
the two peaks reached a minimum count rate of $\approx3330$~count~s$^{-1}$
in the full \nicer band.

We searched for burst oscillations between
612 and 626 Hz with a resolution of 1/8192 s in the 0.5--8.5~keV data
starting from 20 s prior to burst onset, in sliding windows of $T=2$, 4
and 8 s striding at a pace of $T/2$. 
No burst oscillations were observed near
either peak of the X-ray burst to an upper limit of 8\% fractional
amplitude.

\begin{figure}
\centering
\includegraphics[height=3.2in, width=2.55in, angle=-90]{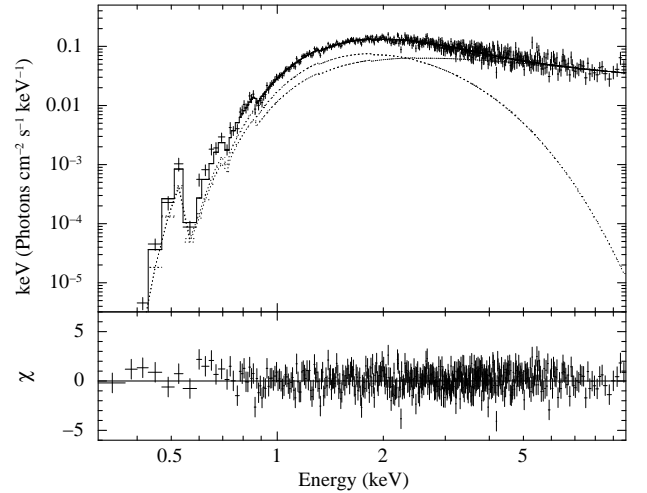}
\caption{
 \nicer  spectrum from the persistent emission prior to the burst. In the 
top panel the 0.3--10 keV energy spectrum is  well described  
by an absorbed  disk-blackbody plus a power-law model. Spectral residuals 
corresponding to the best fitting model are shown in the bottom panel.}
\label{spec-per}
\end{figure}

\begin{figure}
\centering
\includegraphics[height=3.3in, width=2.5in, angle=-90]{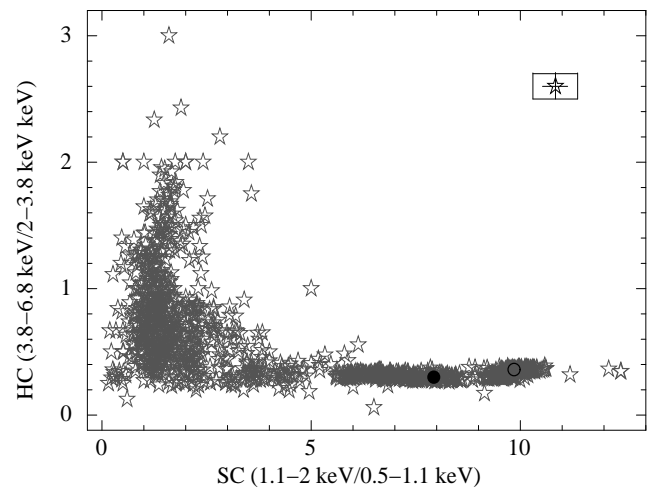}
\caption{ \nicer color-color diagram of \source  observed between 2017 June 
and 2019 April.  The soft color (SC) is defined as the ratio of count rates  
in the (1.1--2.0)/(0.5--1.1)~keV energy bands, whereas the hard color (HC) 
is from the ratio of count rates in the (1.1--2.0)/(0.5--1.1)~keV energy bands. Each 
point indicates a binning time of 128~s with a typical error bars as shown in the right 
corner of the figure. The position of the two bursts observed by \nicer are indicated by 
circles in the banana branch. Thus the source spectral state was soft during the present
double-peaked burst  (solid circle). } 
\label{color}
\end{figure}

\subsection{Persistent Emission} \label{sec:persistent}

The burst considered in this  paper occurred 210~s into the third 
GTI of ObsID 1050070103. We accumulated the first 165~s of good data 
prior to the onset of the burst for the pre-burst emission. The energy 
spectrum extracted from this interval was fitted with a disk blackbody 
({\tt diskbb}) model \citep{Mitsuda1984} 
along with a power-law component. The full model 
{\tt TBabs$\times$(diskbb + power-law)} is able to describe the 0.3--10~keV 
persistent spectrum reasonably well (Figure~\ref{spec-per}).  The goodness of fit per degree of 
freedom is found to be $\chi^2$/$\nu$ = $\chi^2_\nu=1.05$ for  
$\nu=468$ degrees of freedom. The interstellar 
medium absorption $N_{\rm H}$ is described by {\tt TBabs} \citep{Wilms2000}. We 
found  a column density of (0.98$\pm$0.03)$\times$10$^{22}$~cm$^{-2}$ 
with \nicer, which is well within the 1$\sigma$ uncertainty reported 
by \citet{Keek2008} and \citet{Ozel2016}. We do not detect any Fe line 
feature in the pre-burst continuum. 

The spectral parameters of our best-fit model and their 1$\sigma$ 
errors  are: an inner disk temperature $T_{\rm in}$ of the disk black 
body of k$T_{\rm in}=0.65\pm0.03$~keV, an inner disk radius $R_{\rm in}$ of the disk
black body given by $(R_{\rm in}/D_{\rm 10~kpc})^2~{\rm cos}{\rm \theta}=193\pm25$~km$^2$, 
where $D_{\rm 10~kpc}$ is the distance to
the source in units of 10 kpc, a photon index of the power law of
$\Gamma$=1.6$\pm$0.2 and a normalization of the power law at 1 keV of
$0.14\pm0.04$~phot~s$^{-1}$~keV$^{-1}$~cm$^{-2}$.
We used the 
{\tt cflux} model to compute the unabsorbed flux in the 0.3--10 keV band, which
was found to be (1.75$\pm$0.02)$\times$10$^{-9}$~erg~s$^{-1}$~cm$^{-2}$.
By extrapolating beyond the \nicer energy range, the unabsorbed 0.1--100 keV 
band bolometric flux was estimated to be (2.4$\pm$0.1)$\times$10$^{-9}$~erg~s$^{-1}$~cm$^{-2}$. 
We quote only unabsorbed fluxes in this paper.
 At this flux, \source was accreting at a persistent level of 
$\approx$1.6\% of Eddington luminosity. This is calculated with respect to 
the maximum flux (1.5$\times$10$^{-7}$~erg~s$^{-1}$~cm$^{-2}$; \citealt{Galloway2008}) 
observed by \rxte as the Eddington limit.

 We attempted to determine the spectral state before the burst using different methods 
as follows. First, a timing approach was adopted on the \nicer data \citep{van2003}. This was done 
by comparing the source power spectrum, energy spectrum, and bolometric luminosity of the persistent 
emission with archival \rxte observations of \source \citep{van2003}. The analysis suggested that the 
source was possibly in the intermediate lower-left banana branch at the time of the burst. We also 
quantified the spectral state of the  burst in the color-color diagram as 
shown in Figure~\ref{color}. This diagram is obtained by using the available \nicer observations  between 
2017 June and 2019 April at various accretion states. In our analysis, the soft color (SC)
is defined as the ratio of count rates in the (1.1--2.0)/(0.5--1.1)~keV energy bands, while the hard color (HC)
is obtained by the ratio of count rates in the (3.8--6.8)/(2.0--3.8)~keV energy bands 
(see, e.g. \citealt{Bult2018}). Based on the colors prior to the burst (solid circle 
in Figure~\ref{color}), the double-peaked event seems to have occurred in the 
lower banana branch of this atoll source.

\begin{figure}
\centering
\includegraphics[height=3.3in, width=3.2in, angle=-90]{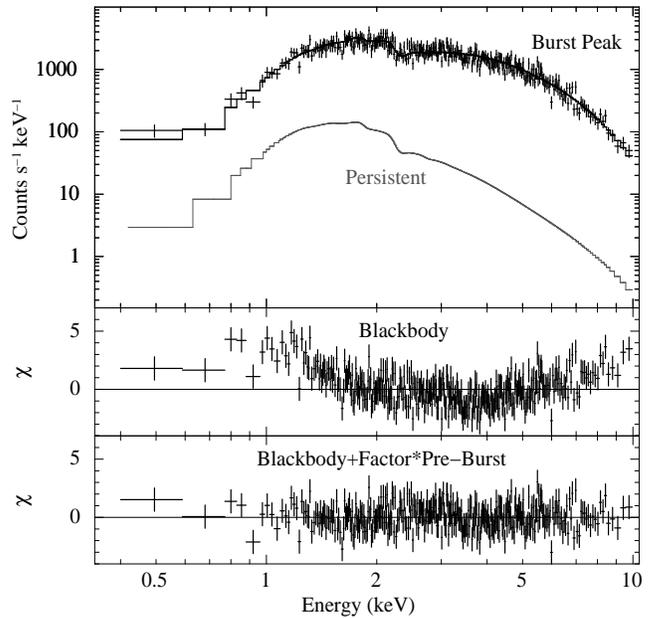}
\caption{The 0.3--10 keV \nicer spectrum obtained from a 1~s time interval  
at the burst peak. The best-fitting model, shown in the top panel, comprises an absorbed blackbody along
with scaled pre-burst (persistent) emission. The middle panel
shows the residuals corresponding to a simple blackbody model
after subtracting the pre-burst emission, while the bottom panel shows the 
residuals for the best fitting $f_a$ model. See Section~\ref{sec:trs} for details.}
\label{spec}
\end{figure}

\begin{figure}
\centering
\includegraphics[height=3.32in, width=4.95in, angle=-90]{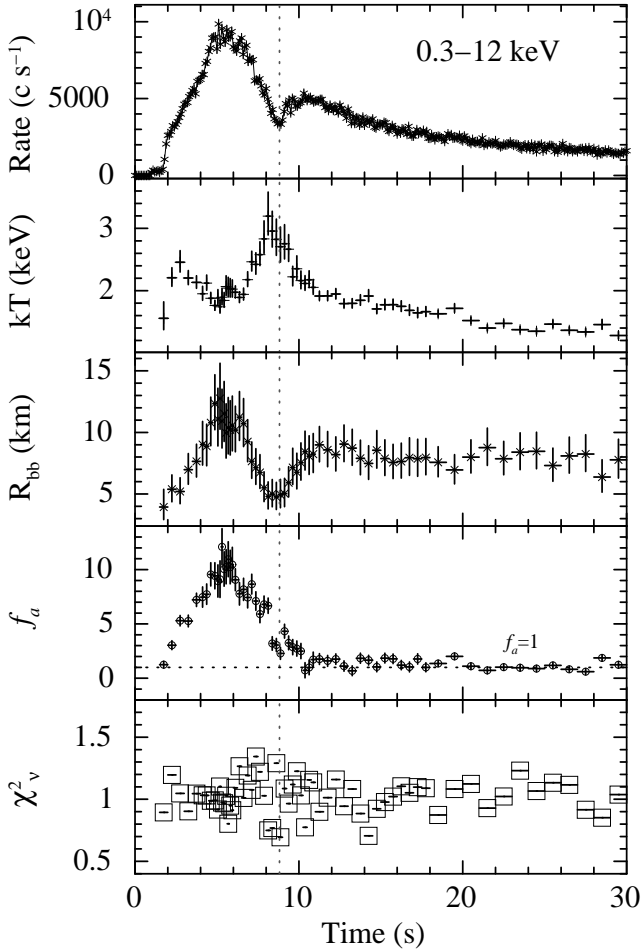}
\caption{Evolution of spectral parameters obtained from burst time-resolved 
spectroscopy. The top panel shows the burst light curve at 0.05~s time resolution. 
The vertical dotted line marks the minimum of the dip feature. The second, third, 
fourth and fifth panels show the temperature, blackbody radius for a distance 
of 4~kpc, scale factor $f_a$ and reduced-$\chi^2$, respectively.  
The horizontal dotted line in the fourth panel is marked at the unity.} 
\label{spec-trs}
\end{figure}

\begin{figure}
\centering
\includegraphics[height=3.42in, width=2.95in, angle=-90]{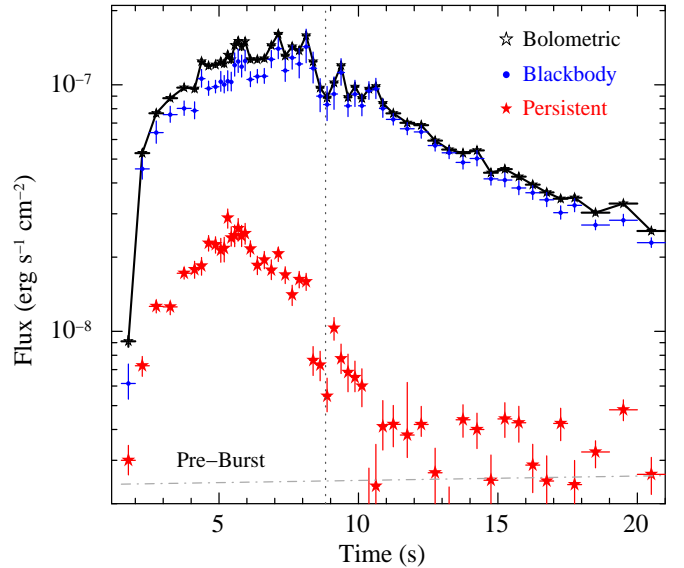}
\caption{Evolution of the 0.1-100 keV bolometric, blackbody, and persistent fluxes during 
the burst. The vertical line shows the time when the dip occurs in the light curve, while 
the horizontal line marks the pre-burst flux level.} 
\label{flux}
\end{figure}

 \subsection{Time-resolved Spectroscopy of the Burst} \label{sec:trs}   
    
To investigate the temporal evolution of the burst spectrum, we 
first divide the burst light curve into three broad time segments:
we use a 1 s bin on the first peak; a 2 s bin on the second peak; 
and a 5 s bin in the tail of the X-ray burst (Figure~\ref{lc}). The 0.3--10 keV spectra 
from these three segments were modelled with a blackbody ({\tt bbodyrad} 
in $XSPEC$) component after subtracting the pre-burst emission as 
a background component. 
The column density for interstellar absorption 
was kept fixed at the value obtained for the persistent emission 
(section~\ref{sec:persistent}). We noticed that this model is not 
sufficient to adequately describe the continuum, especially for the 
burst peak where strong excesses are seen  at both ends of the band 
pass (second panel of Figure~\ref{spec}). The corresponding goodness 
of fit $\chi^2$($\nu$) was found to be 830(506), 633(552) and 
622(577) for the first, second and third time-segments, respectively. 

A better description of the burst emission was obtained by using the
variable persistent flux method \citep{Worpel2013}. For this method, 
we used a blackbody component in addition to the fixed pre-burst spectral 
model obtained from the analysis of the persistent emission (section~\ref{sec:persistent}), 
together with  a free multiplicative factor $f_a$ in the following way: 
{\tt TBabs$\times$(bbodyrad+$f_a\times$(diskbb + power-law)}. The scale factor
$f_a$ accounts for variation in the persistent continuum level with respect to 
the pre-burst value.  However, we note that the burst emission, 
in general, can deviate from pure blackbody radiation due to the 
effect of the neutron star's atmosphere and its fast rotation 
(see, e.g., \citealt{Suleimanov2012}). Thus, the variation in 
$f_a$ likely represents a net outcome from the atmosphere as well as 
a possible contribution from the varying accretion flow during the burst. Our 
current understanding hardly allows us to segregate these effects from the spectrum 
due to degeneracy in theoretical modeling (\citealt{Worpel2015, Degenaar2018}). 
Using the above method, we obtain an improved  fit  with $\chi^2$($\nu$) = 
577(505),  624(551) and 616(576) for the  first, second and third 
time-segments, respectively. 
The residuals corresponding 
to the burst peak interval are shown in the third panel of Figure~\ref{spec}. 
From this preliminary analysis,  we find  variation in  
the blackbody temperature, normalization, and $f_a$ across the burst, 
motivating a more detailed analysis.

In addition to the above spectral modeling, various absorption components 
such as partial covering (\texttt{pcfabs}), absorption through warm 
(\texttt{wndabs}), neutral, and partially ionized materials 
(\texttt{zxipcf}) were also applied in the variable persistent flux, especially 
on a spectrum extracted from a 0.7~s interval at the dip phase,
in order to constrain possible  
origins of the dip in an obscuring medium present close to the 
neutron star surface. None of these models fitted significantly better 
than the $f_a$ model with interstellar absorption with $N_{\rm H}$ fixed 
to 0.98$\times10^{22}$~cm$^{-2}$ as found from the persistent spectrum, 
nor do these models provide evidence for increased absorption due to 
(partially ionized) gas.

Next, we explored time-resolved spectroscopy on a finer time scale in order to 
more fully probe the burst evolution and understand the origin of the two peaks 
(or the one dip) in the profile. For this, we extracted a total of 63  spectra 
with a duration of at least 0.125~s allowing $>$1000 counts per spectrum.  
We fitted all the spectra with the variable persistent flux model
as described above. The results show that the blackbody radius expands 
during the peak of the burst (Figure~\ref{spec-trs}). Considering a 
distance of 4~kpc \citep{Guver2010}, the maximum expansion radius $R_{bb}$ 
is estimated to be 12$\pm$2~km (mean-weighted value from six points on the peak) 
using the blackbody normalization (third panel of Figure~\ref{spec-trs}). 
In contrast to the expansion, the blackbody temperature drops to a value of 
1.83$\pm$0.07 keV after the burst onset (second panel of Figure~\ref{spec-trs}). 
At the same time, the bolometric flux reaches, during one second, a plateau consistent 
with the Eddington flux, at (1.4$\pm$0.2)$\times$10$^{-7}$~erg~s$^{-1}$~cm$^{-2}$ 
for the peak (Figure~\ref{flux}). The measured flux is in the same range as 
the brightest PRE events seen from \source{} with {\it RXTE} \citep{Galloway2008}. 
        
A gradual increase in blackbody temperature was noticed after the photospheric expansion  
(second panel of Figure~\ref{spec-trs}). The temperature reaches $\approx3.2\pm$0.4~keV
during this phase. We found that the dip observed in the light curve does 
not coincide with the maximum temperature, but appears late by about 0.75~s. 
Given this time shift, we suggest that the observed dip is unlikely to be 
related to PRE or any limitation of the instrument passband. A final cooling trend 
is observed $\approx8$~s after the burst onset.

An unusual drop in bolometric flux is detected at the dip
with a significance level of $\approx$3.5$\sigma$ (Figure~\ref{flux}).
This dip reaches a flux value of (8.8$\pm$0.3)$\times$10$^{-8}$~erg~s$^{-1}$~cm$^{-2}$.
Such a drop can also be seen in the evolving 
persistent level. It is interesting to point out that the bolometric 
flux re-brightens $\sim$1~s earlier than the second intensity peak in 
the light curve.  The scaling factor $f_a$ shows a noticeable variation, 
reaching up to a value of 13 during the first peak, and returning to 
unity within nine seconds after onset as shown in Figure~\ref{spec-trs}.

\section{Discussion and Conclusions} \label{sec:dis}

In this paper, we discuss the results of a strong, double-peaked 
burst observed from \source 
using \nicer. Secondary peaks in the burst decay were  detected in 
the soft X-ray light curves, as well as in the bolometric flux. 
Correspondingly, a dip between the first and second peaks is  also 
visible in our study. It is worth noting that this dip shows 
a clear offset with the highest blackbody temperature in the burst tail, 
excluding the possibility of the feature being of instrumental origin, as discussed 
in Section~\ref{sec:intro}. To date, a number of strong PRE bursts 
have been recorded by \nicer in sources like, e.g., 4U 1820--30. In these cases 
the 0.3--12~keV burst profile is singly-peaked despite a maximum color 
temperature of $\approx$4~keV appears in the cooling tail (e.g., \citealt{Keek1820}). 
Based on the above analogy, we argue that the present double-peaked 
burst from \source is of astrophysical origin.

A similar double-peaked burst from \source in the low state was seen 
by \exo in the 1.4--20~keV band (Figure~1 of \citealt{Penninx1989}). It was a 30~s 
long event with a peak intensity 1.35$\times$10$^{-7}$~erg~s$^{-1}$~cm$^{-2}$, 
similar to the present burst. The second peak of the \exo{} burst was detected 
$\simeq$3~s after the first peak, while with the \nicer burst 
the re-brightening occurred about 5~s after the first peak. 
Moreover, a bolometric flux dip was also clearly found in the \exo burst. 
\citet{Penninx1989} explained the double-peaked burst by considering 
multiple generations or release of thermonuclear energy. They also 
considered the possibility of absorption and scattering from an accretion 
disk corona that could have produced a dip in the burst profile.  
Similar bursts with a flux drop have also been observed with \rxte 
(e.g. burst number 12 and 17 in Figure~1 of \citealt{Poutanen2014}), 
establishing the fact that double-peaked bursts are 
occasionally seen in \source irrespective of instrumentation.

Given the similarity, we have examined the relevant hypotheses above
as potential explanations for the double-peaked burst observed with \nicer. 
The effects of transient absorption through a disk corona, hot medium or 
a variable spreading layer  (\citealt{Penninx1989, Kajava2017})  
were explored with detailed spectral analysis. We did not find any evidence 
of additional absorbers at the dip intervals. Thus, absorption/scattering 
of X-ray photons is not a satisfying solution for the observed dip. 
We suggest instead that the peak following the dip in flux is due to 
enhanced emission in the cooling tail.

The thermonuclear flame spreading model of non-PRE bursts can explain 
the origin of a double-peaked burst \citep{Bhattacharyya2006}. According 
to this model, the burning starts at high latitude on the stellar surface 
and propagates toward the equator. When the flame reaches 
the equator, it stalls for a few seconds before spreading into the other hemisphere. 
The stall allows the stellar surface to cool down, causing the observed burst flux to 
temporarily decrease. After a few seconds, the flame continues to spread 
over the remaining surface, producing a secondary rise in flux. While 
this model describes the phenomenological shape of the burst light curve, 
it is unclear what physical mechanism would cause the burning front to stall. 
A potential explanation may be related to the interaction between the burning 
front and the spreading flow of accreted matter \citep{Inogamov1999, 
Bhattacharyya2006}.

Alternatively, re-burning of fresh or leftover material 
(see, e.g., \citealt{Keek2017} and references therein) may produce 
the second  peak in the cooling tail of a burst. It is not clear how the fresh 
material can be kept aside without mixing with burnt fuel \citep{Spitkovsky2002}; 
however, it has been suggested that a hydrodynamical shear instability 
induced by convection during the thermonuclear explosion could lead 
to accumulation of fresh fuel above the burnt material \citep{Fujimoto1988}. 
A model based on nuclear waiting points in the rp-process can also explain 
the double-peaked structure, for accretion rates of a few percents of 
$\dot{M}_{\rm Edd}$, as is the case here \citep{Fisker2004}.

Considering the relatively limited PRE of the present burst, 
it seems that only a part of the neutron star surface is involved 
during the first peak of the burst. Strong convective mixing does 
likely occur during this peak, which eventually leads, after touch-down, 
to the ignition of the unburned material, and thus a second brightening.
In a more exotic interpretation one might presume the double 
peaks being the result of two bursts occurring nearly simultaneously on the 
stellar surface. However, we can rule out this model because 
matter needs to be confined to a small region which is only possible in the 
case of magnetized neutron stars, with field strengths $\ge$10$^9$~G 
(\citealt{Cavecchi2011} and reference therein). For the given magnetic field  
(0.5--1.6)$\times$10$^8$~G of \source{} \citep{Asai2013}, the flame should 
easily spread out and produce a single-peak burst profile.

In summary, we have discussed plausible scenarios to explain the 
double-peaked burst   from the source \source.
 The low-energy capability of \nicer enables us for the first time to rule out absorption 
effects as the origin of the dip, as proposed earlier.     
The possibility of shear instability, thermonuclear flame spreading, 
or nuclear waiting points applicable to non-PRE bursts can fit the 
picture. We also favor the scenario of additional 
burning in the cooling tail of the burst, considering the temperature 
evolution across the burst. The re-burning would be feasible only if 
residual or fresh material lies above the cold fuel as a result of 
hydrodynamical instabilities.

It is interesting to note that the scaling factor ($f_a$) goes  
down at the time of the dip. It thus appears that reprocessing of the burst 
emission by the accretion disk halts temporarily  at this phase. 
If the inner disk is somehow briefly affected during the PRE process 
(perhaps due to Poynting-Robertson drag), a reduction of the reflected 
burst flux would also lower the observed flux \citep{Fragile2018}. Nonetheless, 
we can rule out this possibility as the dip occurs a few seconds after the PRE phase. 
A substantial fraction of the observed burst flux is also expected to be 
scattered off the inner disk that could produce the secondary peak 
(see, \citealt{Lapidus1985, He2016} and references therein). 
However, this idea may be discarded because the accretion proceeds 
forward during the burst as shown by the increasing $f_a$.

~\\

\acknowledgments
We sincerely thank the referee for valuable suggestions on the paper. 
This work was supported by NASA through the \nicer mission and the
Astrophysics Explorers Program, and made use of data and software 
provided by the High Energy Astrophysics Science Archive Research Center 
(HEASARC).  This project has received funding from the European Union's 
Horizon 2020 research and innovation programme under the Marie 
Sk{\l}odowska-Curie grant agreement No.\ 713683. This work benefited 
from events supported by the National Science Foundation under Grant 
No. PHY-1430152 (JINA Center for the Evolution of the Elements).
D.A. acknowledges support from the Royal Society.

\facilities{ADS, HEASARC, \nicer}
\software{\textsc{HEAsoft} (v6.24), \textsc{XSPEC} (v12.10.0;  \citealt{Arnaud1996})}

\bibliographystyle{fancyapj}
\bibliography{references}

\end{document}